\documentclass{mn2e}
\usepackage{psfig}
\usepackage{mnras_cite}
\newcommand{\dd}{{\rm d}}

\begin{document}

\title[Entropy Injection as a Global Feedback Mechanism]{Entropy
  Injection as a Global Feedback Mechanism}  
\author[Oh \& Benson] 
{S. Peng Oh \& Andrew J. Benson\\
Theoretical Astrophysics, Mail Code 130-33, Caltech, Pasadena, CA 91125, USA}

\maketitle

\begin{abstract}
Both preheating of the intergalactic medium and radiative cooling of
low entropy gas have been proposed to explain the deviation from
self-similarity in the cluster $L_{\rm X}-T_{\rm X}$ relation and the
observed entropy floor in these systems. However, severe overcooling
of gas in groups is necessary for radiative cooling alone to explain
the observations. Non-gravitational entropy injection must therefore
still be important in these systems. We point out that on scales of
groups and below, gas heated to the required entropy floor cannot cool
in a Hubble time, regardless of its subsequent adiabatic
compression. Preheating therefore shuts off the gas supply to
galaxies, and should be an important global feedback mechanism for
galaxy formation. Constraints on global gas cooling can be placed from
the joint evolution of the comoving star formation rate and neutral
gas density. Preheating at high redshift can be ruled out; however the
data does not rule out passive gas consumption without inflow since $z
\sim 2$. Since for preheated gas $t_{\rm cool} > t_{\rm dyn}$, we
speculate that preheating could play a role in determining the Hubble
sequence: at a given mass scale, high $\sigma$ peaks in the density
field collapse early to form ellipticals, while low $\sigma$ peaks
collapse late and quiescently accrete preheated gas to form
spirals. The entropy produced by large scale shock-heating of the
intergalatic medium is significant only at late times, $z<1$, and
cannot produce these effects.
\end{abstract}

\section{Introduction}

Numerical simulations of dark matter halos reveal an almost universal
density profile over a wide range of masses
\cite{NFW,ghigna00,klyp01,power02}. If the gas faithfully traced the
dark matter, we would expect the bolometric X-ray luminosity to scale
with the cluster virial temperature as $L_{\rm X} \propto T^{2}$
\cite{kaiser86}, while observations indicate significant steepening of
this relation for low-temperature clusters \cite[and references
therein]{pon00}, suggesting that some non-gravitational process is
affecting the properties of the gas. Indeed, observations of the X-ray
surface brightness profiles in poor groups reveal the existence of an
entropy floor, $S=T/n_{\rm e}^{2/3} \sim 100 {\rm keV cm^2}$
\cite{ponmanetal}, in the gas. Such an entropy floor reduces the
central gas density and breaks the self-similarity of gas density
profiles. There are two schools of thought as to how this entropy
floor could have arisen: (i) as a result of non-gravitational heating
of order $\sim 1$keV per particle, arising from winds driven either by
supernovae or active galactic nuclei (e.g.,
\pcite{kaiser,ponmanetal}), (ii) as a result of radiative cooling of
the low entropy gas, leaving behind only the high entropy gas (which
cannot cool within a Hubble time; e.g., \pcite{voitbryan}). Numerical
simulations which incorporate either of these schemes can reproduce
the observed cluster $L_{\rm X}-T$ relation.

In this paper, we critique the first scenario, on the basis of the
disastrous effects the required level of preheating would have on
subsequent galaxy formation. Groups and clusters are the most massive
non-linear structures present today; to have a significant effect on
their gas profile, a large amount of energy must be injected ($\sim
1$keV per particle). The virial temperature of galaxies is much
smaller (for an ${\rm L_{*}}$ galaxy, typically $T_{\rm vir} \sim
0.1$keV); galaxies will therefore be unable to accrete gas heated to
such high entropies. Furthermore, we shall show that the gas they do
manage to accrete cannot cool within a Hubble time. The net effect
would be a sharp downturn in the comoving star formation
rate. \scite{momao} have also recently examined the impact of
preheating on galaxy formation; they find that preheating could have a
strong impact on the X-ray luminosity and morphology of
galaxies. However (unlike the present work), they do not consider if
preheating will suppress subsequent star formation. Throughout, we
assume a cosmology where $(\Omega_{\rm m},\Omega_{\Lambda},\Omega_{\rm b}
h^{2},h,\sigma_{8 h^{-1}})=(0.3,0.7,0.019,0.7,0.9)$, consistent with
current observational constraints (e.g. \pcite{cmb,hstkey,smith02}).

\section{The Impact of Preheating on Gas Cooling} 

\subsection{The overcooling problem}

Recently, it has been pointed out in semi-analytic work
\cite{bryan2000,voitbryan,voitetal} and numerical simulations
\cite{lewis,muanwong,Dave2002} that radiative cooling could be
responsible for the observed entropy floor in groups and
clusters. Here we point out that: (i) radiative cooling without
non-gravitational feedback likely results in severe overcooling of gas
at group scales, where the entropy floor is most clearly seen. (ii)
Depending on its magnitude and redshift of injection, an entropy floor
due to preheating can efficiently regulate overcooling in
groups. Thus, some form of preheating is still likely to be
necessary. This does not preclude an important role for radiative
cooling: it reduces the otherwise severe energetic requirements for
preheating, since the lowest entropy gas simply cools, rather than
being heated up to the entropy floor.

Let us calculate the amount of cooled gas in the absence of feedback,
as a function of halo mass. Unlike other semi-analytic models, we also
wish to consider a dynamic model in which gas is drawn inward and
compressed within a cooling flow, rather than a static model in which
all gas is cooled out to some cooling radius. This is so as to
explicitly check that the increased gas densities due to inward flow
do not significantly increase cooling rates and thus the cooled mass
fraction. We therefore perform 1D spherically symmetric calculations
very similar to that of \scite{tozzinorman}, who compute a sequence of
adiabats generated by the accretion shock and follow their subsequent
evolution due to radiative cooling; refer to \scite{tozzinorman} for
more details on such models. This is an implicitly Lagrangian scheme
and allows us to directly track the entropy evolution of each mass
shell. The main difference between our treatment and theirs is that we
only implement gas cooling within a static potential, rather than an
evolving potential; experimentation has found that this makes little
difference. We also consider the dark matter to impose a fixed
potential, and ignore the self-gravity of the gas.

We approach the problem by considering the sequence of adiabats
through which a gas shell evolves. The initial entropy profile of the
gas as a function of the mass profile, $K(M) \equiv P(M)/\rho
(M)^{\gamma}= k_{\rm B} T/\mu m_{\rm p} \rho^{2/3}$, can be determined by
using the shock jump conditions if the temperature and density at the
accretion shock are known. We assume the dark matter distribution
follows an NFW profile (Navarro, Frenk, \& White 1997). We assume a
perfect gas, $P= \frac{\rho}{\mu m_{\rm H}}k_{\rm B} T$, and a polytropic
equation of state, $P= K \rho^{\gamma}$, where $\gamma=5/3$.
\scite{tozzinorman} find that the average growth of the main
progenitor of the dark matter halo, found by running Monte Carlo
realizations based on the extended Press-Schechter formula (e.g.,
Lacey \& Cole 1993), can be approximated by the formula: $m(z)=
\left( \frac{1+z}{1+z_{\rm o}} \right)^{-\left[B+A{\rm log} \left(
\frac{1+z}{1+z_{\rm o}} \right) \right]}$, where $A$ and $B$ depend on
cosmology, the final mass $M_{\rm o}$, and the observed redshift $z_{\rm o}$
(we consider $z_{\rm o}=0$ in this paper). As a function of $(M_{\rm DM},z)$
the infall velocity is then found from:
\begin{equation}
\frac{v_{\rm i}^{2}}{2}=\frac{v_{\rm ff}^{2}}{2} + \Delta W -
\frac{c_{\rm s}^{2}}{\gamma-1} \left[ \left( \frac{\rho_{\rm ta}}{\rho_{\rm e}}
\right)^{\gamma-1} -1 \right] ,
\end{equation}
where $\frac{v_{\rm ff}^{2}}{2}= \frac{GM}{R_{\rm s}} - \frac{G M}{R_{\rm ta}}$
($R_{\rm s} \approx R_{\rm vir}$ is the shock radius, and $R_{\rm ta} \approx 2
R_{\rm vir}$ is the turnaround radius), $\rho_{\rm e}$ is the external
preshock density at $R=R_{\rm s}$, and $c_{\rm s}= (\gamma K_{\rm IGM}
\rho_{\rm e}^{\gamma-1} )^{1/2}$ is the sound speed where $K_{\rm IGM}$ is the
preshock entropy of the IGM gas being accreted. The postshock
temperature is then given by (Landau \& Lifshitz 1957; Cavaliere,
Menci \& Tozzi 1998):
\begin{equation}
k_{\rm B} T_{\rm i} = \frac{\mu m_{\rm p} v_{\rm i}^{2}}{3} \left[ \frac{ (1+
\sqrt{1+\epsilon})^{2}}{4} + \frac{7}{10}\epsilon - \frac{3}{20}
\frac{\epsilon^{2}}{(1+
\sqrt{1+\epsilon})^{2}} \right] ,
\label{PS_temp}
\end{equation}
where $\epsilon \equiv 15 k_{\rm B} T_{\rm e}/ 4 \mu m_{\rm p} v_{\rm i}^{2}$. The postshock density is given by $\rho_{\rm i}=g \rho_{\rm e}$, where
the shock compression factor $g$ is given by (Cavaliere, Menci, \&
Tozzi 1997):
\begin{equation}
g= 2 \left( 1 - \frac{T_{\rm e}}{T_{\rm i}} \right) + \left[ 4 \left( 1 -
\frac{T_{\rm e}}{T_{\rm i}} \right)^{2} + \frac{T_{\rm e}}{T_{\rm i}} \right]^{1/2}.
\end{equation}
From this one can compute the postshock adiabat 
\begin{equation}
K_{\rm shock}= k_{\rm B}
T_{\rm i}/\mu m_{\rm p} \rho_{\rm i}^{\gamma-1}.
\label{eqn:Kshock}
\end{equation} 

We have also computed the entropy profiles assuming that the gas
traces the NFW profile and is in hydrostatic equilibrium, a model used
by \scite{voitetal}. We have found that the entropy profile computed
with either model is very similar; both are sharply increasing
functions of radius. We have also found that the entropy profile for
different cluster masses (scaled to the entropy of the outermost
shell) is fairly self-similar across different cluster masses. This is
not surprising, since density and temperature profiles are fairly
self-similar in these models; they only differ to the extent that
their central concentration (which depends primarily on the collapse
redshift) varies. We will examine the effects of an entropy floor in
these models in \S \ref{subsection:gas_cool}.

Given an initial entropy profile $K(M,t=0)$, we simultaneously solve
the equations for the variables $K(M,t),\rho(M,t),R(M,t)$ (which
automatically also yields $T(M,t)=K(M,t) \rho(M,t)^{\gamma-1} \mu
m_{\rm H} /k_{\rm B}$):
\begin{eqnarray}
\frac{\dd P}{\dd M_{\rm gas}}= - \frac{G M_{\rm DM}}{4 \pi r^{4}} \nonumber \\ 
\frac{\dd r}{\dd M_{\rm gas}}= \frac{1}{4 \pi \rho_{\rm gas} r^{2}} 
\label{DE_system}
\end{eqnarray}
Given the initial entropy profile we have calculated, we can compute
  the initial pressure profile of the gas, if we assume the boundary
  condition $\rho(R_{\rm vir}) = f_{\rm B} \rho_{\rm NFW}(R_{\rm
  vir})$, where $f_{\rm B}=\Omega_{\rm b}/\Omega_{\rm m}$ is the
  universal baryon fraction. The pressure and entropy profile fully
  specify the initial density and temperature profile of the gas. We
  now allow for radiative cooling to operate, and update the entropy
  profile via:
\begin{equation}
\frac{\dd}{\dd t}{\rm ln (K)}= -\frac{1}{\tau_{\rm cool}(K)},
\end{equation}
where the cooling time is: 
\begin{equation}
\tau_{\rm cool} \equiv \frac{3}{2} \frac{k_{\rm B} T}{\Lambda_{\rm N}}
\frac{\rho_{\rm gas}}{\mu m_{\rm p}} ,
\end{equation} 
and we use a fit given by \scite{tozzinorman} to the cooling function
$\Lambda_{\rm N}$ of Sutherland \& Dopita (1993), which includes both
free-free and line emission, for a metallicity of $Z=0.3
Z_{\odot}$. We choose our timestep so that it corresponds to the
cooling time of the third innermost shell; this is sufficient to
obtain convergent results without excessively long computation
times. As gas cools and drops out at the center, an inward cooling
flow develops. After each timestep, we recompute hydrostatic
equilibrium in equation (\ref{DE_system}) according to the new profile
of adiabats $K(M)$, updating the density and temperature profile of
the gas. Any gas which has cooled below $T \sim 5 \times 10^{5}$K is
removed from the calculation and placed in a cold phase at the center
of the cluster; we track $M_{\rm cool}$ and $M_{\rm hot}=M_{\rm
gas,0}-M_{\rm cool}$. We do not consider distributed mass drop-out in
this calculation. When recomputing hydrostatic equilibrium, we use the
boundary conditions:
\begin{eqnarray}
r(0)&=&0 \\ \nonumber
r(M_{\rm hot})&=&r_{\rm end} \\ \nonumber
\rho(M_{\rm hot})&=&\left[
\rho_{\rm vir}^{\gamma-1} + 
\frac{\gamma-1}{\gamma K_{\rm vir}} \int_{r_{\rm vir}}^{r_{\rm end}} - \frac{G M_{\rm DM}(r)}{r^{2}} \dd r
\right]^{1/(\gamma-1)}
\end{eqnarray}
where $\rho_{\rm vir}=\rho_{\rm o}(r_{\rm vir})$, $K_{\rm vir}=K(r_{\rm vir})$ are the
initial density and entropy of the outermost mass shell. The last
boundary condition uses the fact that the outermost shell is
compressed adiabatically during the inward flow, since for this shell,
$t_{\rm cool} \gg t_{H}$. Note that since we have 3 boundary conditions
for 2 ordinary differential equations, we can solve for $r_{\rm end}$,
which is an eigenvalue of the problem. Since this is a boundary value
problem, we use relaxation to solve the equations. The results of our
calculations agree very well with those of Tozzi \& Norman (2000),
although they use a slightly different set of equations and boundary
conditions. In our calculation, we assumed the dark matter potential
was fixed and merely evolved the equations for a Hubble time, allowing
the gas to cool. We have also experimented with a growing dark matter
potential; the results do not differ significantly.
   
\begin{figure}
\psfig{file=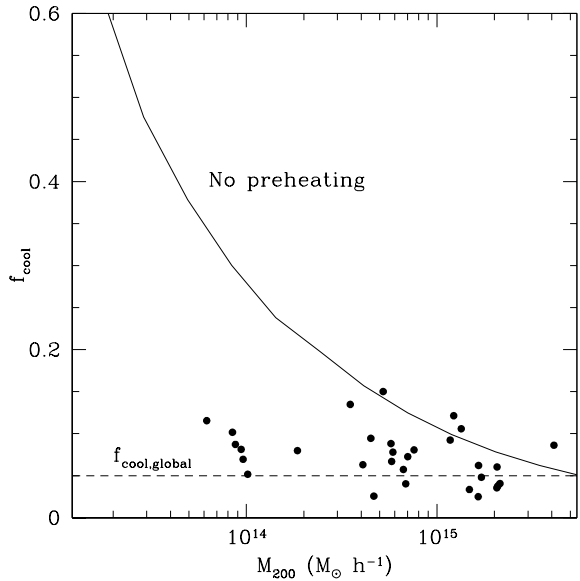,width=80mm}
\caption{The fraction of gas which cools in a Hubble time $f_{\rm
  cool}$ as a function of halo mass, as computed in our models without
  preheating. The points depict the fraction of baryons in the cold
  phase in a sample of clusters from Roussel et al (2000). The dashed
  line shows the best estimate of the global cold baryon fraction from
  the K-band luminosity function of Cole et al (2001). Radiative
  cooling alone without feedback results in severe overcooling,
  particularly in groups.}
\label{fig:Mcool_with_data}
\end{figure}

In Figure \ref{fig:Mcool_with_data}, we show the fraction $f_{\rm cool}$
of gas which cools in a Hubble time in our models without preheating,
computed according to the convention:
\begin{equation}
f_{\rm cool}= \frac{M_{\rm cool}}{M_{\rm tot}}
  \frac{\Omega_{\rm m}}{\Omega_{\rm b}},
\label{eqn:fcool}
\end{equation}
where $M_{\rm cool}$ is the mass in the cool phase and $M_{\rm tot}$
is the total gravitating mass. We prefer this to the other widely used
convention: $f_{\rm cool} = M_{\rm cool}/(M_{\rm cool} + M_{\rm
hot})$, where $M_{\rm hot}$ is the gas mass in the hot phase, because
the total gas mass $M_{\rm hot}+M_{\rm cool}$ in a fixed halo
potential decreases with an increasing entropy floor in preheating
scenarios (a high entropy floor hinders shallow potential wells from
accreting gas). We also choose to show halo mass rather than the X-ray
temperature because for a fixed potential, the emission weighted
temperature increases with the entropy floor (since gas at the cluster
center, which dominates the X-ray emission, becomes hotter). These
conventions allow the clearest depictions of the effect of an entropy
floor on gas cooling. Also shown as points is $f_{\rm cool}$ inferred
from the data of \scite{rousseletal}, who compile a list of total
stellar mass and total gravitating mass for a sample of X-ray
clusters. As in \scite{Baloghetal} (see their paper for a very lucid
discussion of the overcooling problem), we multiply the total stellar
mass by 1.1 to account for gas in atomic and molecular form; we also
show the global cooled gas fraction of $f_{\rm cool} \sim 5\%$
inferred from the K-band luminosity function by \scite{coleetal}. The
latter figure is somewhat lower than the customarily quoted value of
$6\% < f_{\rm cold} < 17\%$ found by \scite{fgp}, but much more robust
as it does not depend on the highly uncertain relative abundances of
galaxies of different morphological types, which have very different
$M/L_{\rm B}$; by contrast $M/L_{\rm K}$ varies by at most a factor of
two over different stellar population histories \cite{belldejong}.

We see that while radiative cooling alone may be consistent with
observations in the most massive clusters, severe overcooling of the
gas would occur on group scales and below if radiative cooling alone
were to operate without feedback, with $\sim 50 \%$ of the gas
entering the cold phase. In our models we have computed the entropy
profiles and allowed for gas inflow and compression in a cooling
flow. We also computed static models which assume that the gas traces
the dark matter and that all gas cools out to some cooling radius
$r_{\rm cool}$ where $t_{\rm cool} \sim t_{H}$
\cite{whitefrenk,voitetal,wuxue}, and obtained very similar
results. We comment on this robustness in \S
\ref{subsection:gas_cool}.

This overcooling effect is in fact consistent with all numerical
simulations which claim to explain the observed $L_{\rm X}-T_{\rm X}$
relation from the effects of radiative cooling alone, although this
has not been sufficiently emphasized. For instance, while
\scite{muanwong} find a global cooled gas fraction of $f_{\rm cool}
\sim 15 \%$, on group scales where $M_{\rm vir} \sim {\rm few} \times
10^{13} {\rm M_{\odot}} h^{-1}$, the cooled mass fraction is $f_{\rm
cool} \sim 30-50 \%$ (see their Figure 3).  \scite{Dave2002} find that
while $f_{\rm cool} \sim 24 \%$ globally in their simulations, for
halos with $\sigma_{\rm 1D} \sim 100 {\rm km \, s^{-1}}$, then $f_{\rm
cool} \sim 80 \%$, while for halos with $\sigma_{\rm 1D} \sim 500 {\rm
km \, s^{-1}}$, then $f_{\rm cool} \sim 50\%$; for these extreme
levels of gas cooling, consistency with the observed $L_{\rm X}-T_{\rm
X}$ relation can be achieved. The overcooling problem is often dealt
with in numerical simulations by deliberately restricting the
resolution to $\sim 10^{11} M_{\odot}$; all simulations of higher
resolution generically suffer from overcooling \cite{Baloghetal}, as
was pointed out by \scite{suginoharaostriker}. Based on the
\scite{mulchaeyetal} sample, \scite{bryan2000} claims that the cool
gas fraction increases from cluster to groups scales, and consistency
with the radiative cooling model can be achieved. However, he computed
$f_{\rm cool}=M_{\rm cool}/(M_{\rm cool}+M_{\rm hot})$, and estimates
of the hot gas component are likely to be systematically biased
downwards in groups due to the limited radial extent of the X-ray
profile \cite{rousseletal}.

We therefore contend that radiative cooling alone as an explanation
for the observed scaling relations and entropy floor in groups and
clusters is untenable. The problem is most acute in groups: since most
of the gas in groups is at comparatively low entropy, a larger
fraction of the gas has to cool to recover the observed entropy
floor. In the next section we explore how entropy injection can
ameliorate the overcooling problem.

\begin{figure}
\psfig{file=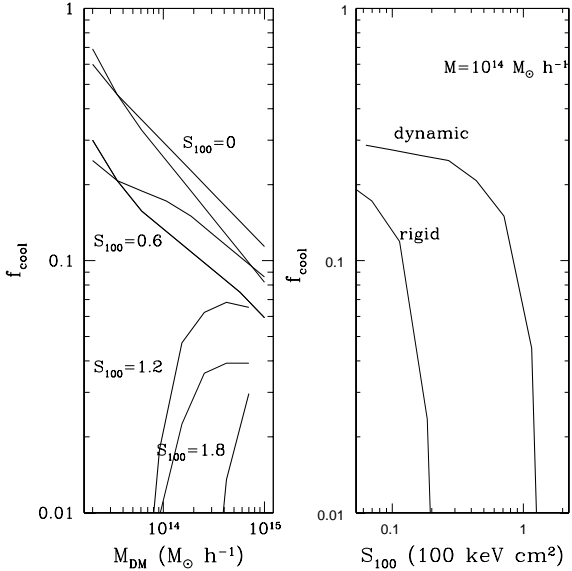,width=80mm}
\caption{The effect of an entropy floor $S_{100}$ on the cool gas
fraction $f_{\rm cool}$ as given by equation \ref{eqn:fcool}. As the
entropy floor increases, the fraction of gas able to cool in a Hubble
time falls drastically. We track the inward flow and compression of
the gas as it cools (dark solid lines); however, ignoring the work
done on the gas by the gravitational potential and assuming it is
static gives very similar results (light solid lines).}
\label{fig:Mcool_vary}
\end{figure}

\subsection{The Effect of an Entropy Floor on Gas Cooling} 
\label{subsection:gas_cool}

Let us now suppose that some form of external preheating takes place,
in which the IGM is boosted to some entropy floor $K_{\rm IGM}$. What
is the effect of this entropy floor on gas cooling? We can easily
incorporate the effects of a pre-existing entropy floor into our
models. If the infall velocity does not exceed the local sound speed,
then the gas is accreted adiabatically and no shock occurs; the gas
entropy $K_{\rm IGM}$ is therefore conserved. If gas infall is
supersonic, then the gas is shocked to the entropy $K_{\rm shock}$
computed in the previous section. \scite{tozzinorman} found that the
transition between the adiabatic accretion and shock heating regime is
very sharp. Thus to a very good approximation we can set $K(M)={\rm
max}(K_{\rm shock}(M),K_{\rm IGM})$; with this new entropy profile we
can compute density and temperature profiles and gas cooling
properties as before. Similar to previous work (e.g.,
\pcite{tozzinorman,voitetal}), we find that an entropy floor creates a
core in the density profile, decreasing the central density and
increasing the central temperature.

Let us define, as is customary, the entropy parameter $S\equiv
T/n_{\rm e}^{2/3}$, and
\begin{equation}
S_{100} \equiv \left( \frac{S}{100 {\rm keV cm^{2}}} \right).
\label{eqn:s100_define}
\end{equation}
For mean molecular weight $\mu=0.6$, this is related to our previously
defined entropy parameter in equation (\ref{eqn:Kshock}) via
$S_{100}=6.27 K_{34}$, where $K_{34}=10^{34} {\rm erg \, cm^{2} \,
g^{-5/3}}$. Typical values of the observed entropy floor range between
$S_{100} \sim 1$ \cite{ponmanetal}, to $S_{100} \sim 4$
\cite{finoetal}. Only values toward the lower end of such estimates
can be produced by radiative cooing alone \cite{voitbryan}, whereas
values toward the higher end are necessary to reproduce the observed
$L_{\rm X}-T_{\rm X}$ relation \cite{tozzinorman}.

In Figure \ref{fig:Mcool_vary}, we show how the cool gas fraction
$f_{\rm cool}$ varies with the entropy floor $S_{100}$. We see two
main features in this plot: (i) an increase in the entropy floor
drastically decreases the cooled gas fraction. An entropy floor thus
has the potential to solve the overcooling problem. (ii) The results
of the calculation where the inward flow and compression of the gas as
it cools is tracked (dark solid lines) is very similar to the case
when this is ignored and the gas is assumed to be static (light solid
lines). The compression of the gas by the gravitational potential as
it flows inwards only causes a mild increase in the final cooled gas
fraction $f_{\rm cool}$; naively, one might have expected that the
increased densities would resulted in much more rapid cooling.

\begin{figure}
\psfig{file=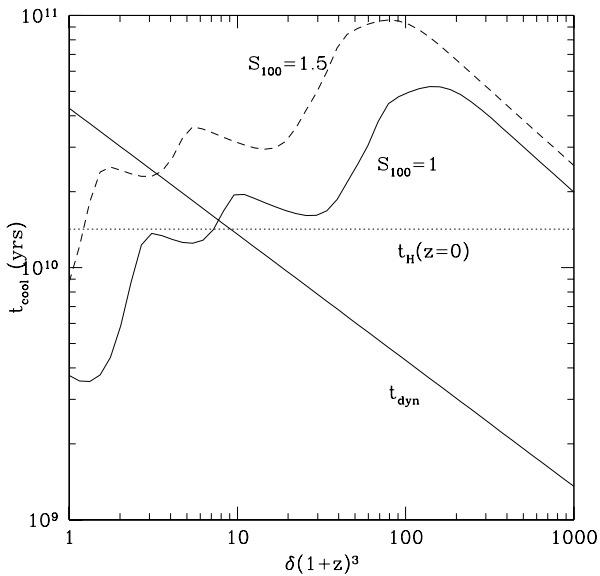,width=80mm}
\caption{The cooling time as a function of density $n=2 \times 10^{-7}
\delta (1+z)^{3} {\rm cm^{-3}}$ for gas at fixed entropy $S_{100}$, as
defined in equation \ref{eqn:s100_define}. Also shown are the Hubble
time at $z=0$ and the dynamical time $t_{\rm dyn} \propto
n^{-1/2}$. The cooling time depends only weakly on density; for the
range of temperatures relevant for galaxy-sized halos, $T < 0.1$keV or
$\delta (1+z)^{3} < 100 S_{100}^{3/2}$, it typical {\it increases}
with density. In general, gas above the entropy floor $S> S_{100}$
cannot cool in a Hubble time.}
\label{fig:tcool_vs_delta}
\end{figure}

\begin{figure}
\psfig{file=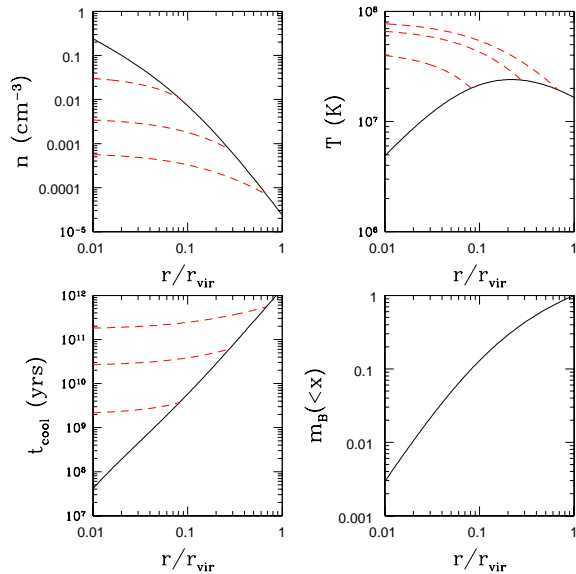,width=80mm}
\caption{Evolution of the density, temperature and cooling time as a
parcel of gas is brought inwards adiabatically in a $M_{\rm
DM}=10^{14} M_{\odot} h^{-1}$ halo. Because gas at large radii is at
high entropy, an inward flow only brings about modest compression; the
rise in temperature offsets the cooling rate increase due to the
increased density, and overall the cooling time only shows a modest
decrease. In fact, for adiabatic compression the cooling time is
almost independent of the density, and depends only on the entropy.}
\label{fig:cooling_times}
\end{figure}

To understand these results, let us consider how gas cooling times
evolve at fixed entropy. If gas with entropy $S_{100}$ evolves
adiabatically, the gas temperature depends on its density as:
\begin{equation}
T = 4.7 \times 10^{4} S_{100} \delta^{2/3} (1+z)^{2} \ {\rm K},
\end{equation}
where $\delta= n_{\rm e}/\bar{n}_{\rm e}$ is the overdensity of the gas. The
isobaric cooling time is then:
\begin{eqnarray}
t_{\rm cool} = \frac{5 n_{\rm e} k_{\rm B} T}{2 \mu n_{\rm e} \Lambda(T)} &\propto&
S^{3/2} {\rm (T < 10^{7} {\rm K})} \\ \nonumber
&\propto& \frac{S^{1/2}}{n^{2/3}} (T>1 10^{7} {\rm K})
\end{eqnarray} 
where $\mu \approx 0.6$ is the mean molecular weight. We have used
$\Lambda(T) \propto T^{1/2}$ for $T > 10^{7}$K (free-free cooling
dominated regime) and $\Lambda(T) \propto T^{-1/2}$ for $10^{5.5} {\rm
T} < T < 10^{7}$K (line cooling dominated regime). We can explicitly
compute the cooling time as a function of density for a fixed adiabat;
the results are shown in Figure \ref{fig:tcool_vs_delta}. We have used
a fit \cite{nath} to the \scite{sutherlanddopita} cooling function,
assuming that $Z=0.3 Z_{\odot}$; the detailed temperature dependence
of the cooling function in the line cooling dominated regime is not
strictly $\Lambda \propto T^{-1/2}$, which is why there is a weak
density dependence in the cooling time at low density. The most
important feature to note is that gas above a certain entropy floor
$S_{\rm crit} > S_{100}$ cannot cool in a Hubble time, regardless of
its density and/or how it is adiabatically compressed. It is thus easy
to understand how an entropy floor drastically decreases the cool gas
fraction $f_{\rm cool}$ as $S_{\rm preheat} \rightarrow S_{\rm crit}$. We can also understand why the compression of gas in an inward cooling flow has
little effect on $f_{\rm cool}$: the cooling rate depends on the
entropy of the gas, not the density. We show this effect in
Fig. \ref{fig:cooling_times}, which shows how the gas density,
temperature and cooling time evolve as a gas parcel at fixed entropy
is brought inwards. Because gas at large radii is at high entropy, an
inward flow only brings about modest compression; the rise in
temperature offsets the cooling rate increase due to the increased
density, and overall the cooling time only shows a very modest
decrease.

The fact that the observed 'entropy floor', $S_{\rm preheat}$,
corresponds to a critical entropy where $t_{\rm cool} > t_{H}$ is
perhaps unsurprising: if $t_{\rm cool}(S_{\rm preheat}) < t_{H}$, then
we might expect radiative cooling to erase all evidence for an entropy
floor. Indeed, it was the central insight of \scite{voitbryan} that
since $t_{\rm cool} (S_{\rm crit}) \sim t_{\rm H}$, the entropy floor
could have arisen from a floor in the radiative cooling time, rather
than an epoch of preheating. However, these statements have very
interesting consequences for gas cooling in smaller halos associated
with $L \le L^{*}$ galaxies. For the temperature regime we are
interested in ($T \le 0.1$keV, or $\delta < 100 S_{100}^{3/2}
(1+z)^{-3}$; gas hotter than this cannot be accreted by galaxies), the
cooling time either stays constant or {\it increases} with density for
isentropic gas, rather than decreasing with density. Thus, regardless
of its final density profile, gas which is accreted adiabatically into
a galaxy-size halo has a fairly constant cooling time which depends
only upon its initial entropy. Gas which ends up bound to halos at low
redshift will typically fulfill $\delta (1+z)^{3} > 10$ at all
times. {\it Thus none of the gas which is heated to the entropy floor
will be able to cool in a Hubble time in galaxies. Once the entropy
floor in the IGM is established, the supply of fresh gas for star
formation is cut off}. This is independent of the details of the
accretion of gas and halo mergers; in general, shocks due to this
processes can only increase the entropy of the gas, not reduce
it. Thus, after $z_{\rm preheat}$ where the IGM is heated up to some
adiabat $S_{\rm crit}$, fresh accretion of gas from the IGM is
halted. Hereafter, we shall follow through on the logical consequences
of the observed entropy floor in groups and clusters, if it arose from
an epoch of preheating, to assess whether such a scenario is realistic
or not.

\subsection{Bounds on the epoch of preheating}

We can use the value of the entropy floor to obtain a lower bound on
the epoch of preheating. The gas must be preheated at a sufficiently
high redshift that it is accreted adiabatically: i.e., $S_{\rm
preheat} > S_{\rm shock}$, where $S_{\rm shock}$ is the entropy
acquired due to the virialization shock in the absence of
preheating. Otherwise, the gas will be shocked to a higher adiabat,
erasing the signature of preheating. The entropy acquired at the
accretion shock $S_{\rm shock}=T/n^{2/3} \propto T_{\rm shock}
(1+z)^{-2}$ where $T_{\rm shock}$ increases with time as the depth of
the potential well grows. Therefore, without preheating the entropy of
freshly accreted gas increased monotonically with time. There is thus
a minimum redshift $z_{\rm preheat}$ below which preheating leaves no
imprint on the entropy profile of a cluster, as the entropy of the gas
is boosted above the entropy floor by the strong virialization
shock. For a present day cluster of mass $M_{\rm o}$, we can compute
its entropy profile if we know its accretion history and mass as a
function of redshift $M(z)$. We can use this to bound the epoch of
preheating: if $S_{\rm preheat} \sim 100 {\rm keV cm^2}$, then for the
median accretion history of groups with $T \sim 1 {\rm keV}$ (where
the entropy floor is most prominent), the epoch of preheating must be
$z \ge 2$, when the group acquired its core.

\begin{figure}
\psfig{file=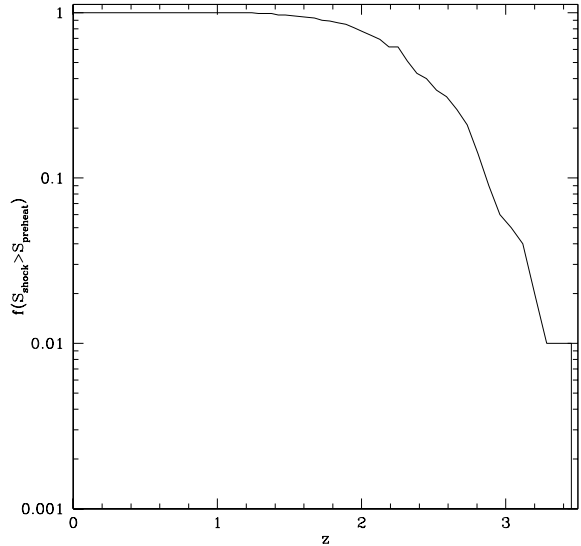,width=80mm}
\caption{The mass fraction of a $10^{14}M_\odot$ halo at $z=0$ in
halos which are unaffected by a preheating entropy of $S_{\rm
preheat}=100$keV cm$^2$ as a function of redshift. In order to affect
a significant fraction of the cluster's mass, preheating must occur
before $z\approx 1$--2.}
\label{fig:fracz}
\end{figure}

A more detailed calculation can be carried out using the extended
Press-Schechter theory. We compute the merging history of several
$10^{14}M_\odot$ halos identified at $z=0$, using the algorithm of
\scite{cole00}, and locate at each redshift the most massive
progenitor. For each progenitor we compute $S_{\rm shock}$, and
identify those progenitors for which $S_{\rm shock}>S_{\rm
preheat}$. These progenitors are unaffected by preheating. Therefore
we are able to estimate the fraction of clusters whose most massive
progenitors are unaffected by preheating as a function of
redshift. This fraction is shown in Figure~\ref{fig:fracz}. At $z=2.2$
approximately 50\% of cluster progenitors are affected by preheating,
while at $z=3$, approximately 95\% of cluster progenitors are
affected. Thus, preheating must occur prior to $z=2$--3 in order to
have a significant effect on the cluster gas.

Similarly, from their sample of 9 groups of galaxies observed with
ASCA and ROSAT, \scite{finoetal} conclude from extended
Press-Schechter theory that the accreted gas reaches an entropy level
of 400 ${\rm keV \, cm^2}$ by $z \sim 2.0-2.5$, while such high
entropies were not present at $z> 2.8-3.5$, implying that preheating
happened in a fairly instantaneous fashion at $z \sim 2-3$.

If the entropy of the IGM was indeed boosted to such high levels at $z
\sim 2$, this implies that the supply of cold gas to galaxies was shut
off at that epoch. Remarkably, in their models of cosmic chemical
evolution, \scite{peifall} concluded that the data was consistent with
no gas inflow since $z\sim 2$. However, their models were computed for
a flat SCDM model, which has since fallen out of favour. By performing
a stripped down version of their models, we can nonetheless ask
whether the data are consistent with no gas inflow since $z_{\rm
preheat}$. If fresh gas cannot cool, then
\begin{equation}
\dot{\Omega}_{\rm g}= -\dot{\Omega}_{*}(1-R),
\end{equation}
where $\dot{\Omega}_{\rm g},\dot{\Omega}_{*}$ are the comoving
densities of cold interstellar gas and stars, all in units of the
present critical density, and $R$ is the recycled fraction, which is
$0.3-0.4$ for typical IMFs. In this case, the comoving gas density at
the epoch of preheating is:
\begin{equation}
\Omega_{\rm g}(z_{\rm preheat}) =\Omega_{\rm g}(0)+ (1-R) \int_{0}^{z_{\rm preheat}}
\dd z \frac{\dd t}{\dd z} \dot{\Omega}_{*}.
\label{eqn:omega_HI}
\end{equation}
In principle this is only a lower limit since it neglect outflows due
to supernova explosions, galactic winds, etc. To model the evolution
of the comoving star formation rate we use a fit to the data from
\scite{pozmadau} for a $\Lambda$CDM cosmology:
\begin{equation}
R_{\rm SFR}(z)=C(z) h_{65} \frac{{\rm exp}(3.4z)}{{\rm exp}(3.4z)+22}
{\rm M_{\odot} yr^{-1} Mpc^{-3}},
\label{eqn:SFR}
\end{equation}
where $C(z)=A h_{65}(\Omega_{\rm m}(1+z)^{3}+\Omega_{\rm
k}(1+z)^{2}+\Omega_{\Lambda})^{1/2}/(1+z)^{3/2}$ is a correction
factor for a $\Lambda$CDM cosmology. We use this fitting function
largely to model the precipitous drop in the comoving star formation
rate at $z \sim 1$; we choose the normalization factor $A$ so that
$(1-R)\int_{0}^{5} \dot{\Omega}_{*} = f_{*} \Omega_{\rm b}= 2 \times
10^{-3}$, where $f_{*}=5\%$ is the total baryon fraction in stars seen
today \cite{coleetal}, as inferred from the K-band luminosity
function. They also present two other models which vary the amount of
star formation at high redshift; however, we shall see that these
variations make little difference to our calculations. To estimate the
comoving density of cold gas, we use estimates of the neutral gas in
damped Ly-alpha systems \cite{perouxetal,raoturnshek}, as well as
estimates of the local HI density from 21 cm emission surveys
\cite{zwaanetal,rosen}.

\begin{figure}
\psfig{file=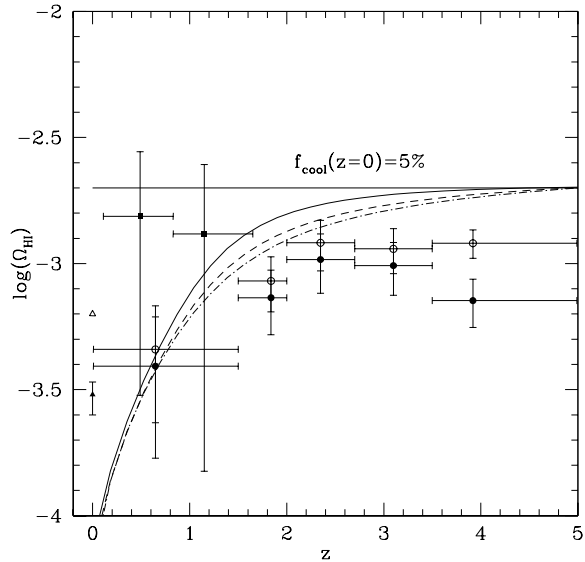,width=80mm}
\caption{The comoving density of interstellar gas required at redshift
$z$, if all gas inflow/cooling ceases at the redshift, in order to be
consistent with the observed comoving star formation rate as a
function of redshift, as computed from equations \ref{eqn:omega_HI}
and \ref{eqn:SFR}. The curves are normalized to produce the total
comoving density of stars seen at z=0, $\Omega_{*}(0) \approx 2 \times
10^{-3}$, as inferred from the K-band luminosity function (Cole et al
2001); any increase in this estimate would move the curves upward. The
points show the inferred comoving neutral gas density in damped
Lyman-alpha systems (Peroux et al 2002; filled circles; open circles
denote their correction for gas in lower column density sytems), Rao
\& Turnshek 2000 (filled squares) and seen locally in 21 cm emission
(Zwaan et al 2001 (filled triangle), Rosenberg \& Schneider 2002 (open
triangle)). In general these represent lower limits. There is
sufficient cold gas at high redshift to be consistent with a shutoff
in gas accretion since $z \sim 2$.}
\label{fig:omega_HI}
\end{figure}

In Figure \ref{fig:omega_HI}, we show the results of computing
$\Omega_{\rm g}(z)$ from equation (\ref{eqn:omega_HI}). This shows the
comoving density of interstellar gas required at redshift $z$, if all
gas inflow/cooling ceases at that redshift, in order to be consistent
with the observed comoving star formation rate as a function of
redshift and the local observed gas density. One simple way to phrase
this constraint is that $\Omega_{\rm g}(z_{\rm preheat}) \approx
\Omega_{\rm g}(0)+\Omega_{*}(0) \approx 2 \times 10^{-3} (f_{\rm
cool}/0.05)$, where $f_{\rm cool}$ is the cool global baryon fraction
today. We see from the figure that there is sufficient cool gas at
high redshift to produce the stars seen today; in particular,
preheating at high redshift $z \sim 2$ can be plausibly accomodated
without subsequent gas cooling and inflow. Note that the normalization
of $\Omega_{\rm g}(z)$ is fixed by the assumed value of $f_{\rm
cool}(z=0)$ (which is uncertain to at least a factor of 2); if this
value increases all the curves shift upward. Also, $\Omega_{\rm DLA}$
is obviously merely a lower bound on $\Omega_{\rm cool}$. The surveys
are based on optically selected quasars, so dusty damped Ly-alpha
systems will not be represented if they dim the background quasar so
that it falls out of a magnitude limited survey; also, a signficant
fraction of the gas could lie in lower column density systems or in
molecular or ionized form. Therefore, despite uncertainties in the
modelling and data, we can at least conclude that the data do not rule
out the absence of gas cooling and inflow into galaxies since $z\sim
2$.

\section{Discussion}

\subsection{Shock heating of the Warm Hot Intergalactic Medium}
Cosmological simulations predict that $\sim 30-40 \%$ of present day
baryons reside in a warm-hot intergalactic medium (WHIM), with
temperatures ${\rm 10^{5} < T < 10^{7} K}$, and mean overdensities
$\delta \sim 10-30$, rather than in virialized objects
\cite{CO99,Croftetal,Daveetal}. Can gas heated in such large scale
shocks accrete and cool in galaxies?  \scite{Daveetal} find that the
WHIM at z=0 approximately follows the equation of state:
$\rho/\rho_{\rm b}= T/10^{4.7}{\rm K}$. We can use this to estimate
the entropy of the WHIM:
\begin{equation}
S_{\rm WHIM}(z=0) \approx 340 \left( \frac{\delta}{30} \right)^{1/3}
{\rm keV \, cm^{-2}}.
\end{equation}
Since $S_{\rm WHIM} > S_{\rm crit}$, none of the gas in the WHIM, even
if accreted onto galaxies, can cool in a Hubble time. Thus, shock
heating by large scale structure could conceivably play a similar role
as preheating in raising the entropy of the IGM and suppressing
accretion and gas cooling.

However, in hierarchical structure formation scenarios, such shock
heating cannot be responsible for the observed entropy floor in groups
and clusters. Shock heating of the WHIM corresponds to $\sim 1 \sigma$
fluctuations turning nonlinear today; the cores of groups and clusters
were assembled at high redshift when they were $ > 2 \sigma$
fluctuations. Gas shocked in the WHIM phase is likely shocked again as
it is boosted to the higher entropies $S \sim 1350 (T/1 {\rm keV})
(n/2 \times 10^{-5}{\rm cm^{-3}})^{-2/3} {\rm keV \, cm^{-2}}$
associated with the outskirts of present-day groups and
clusters. Shock heating of the WHIM is likely only to affect gas
accretion in voids, where the halos have shallow potential wells and
are unable to accrete gas of high entropy.

\begin{figure}
\psfig{file=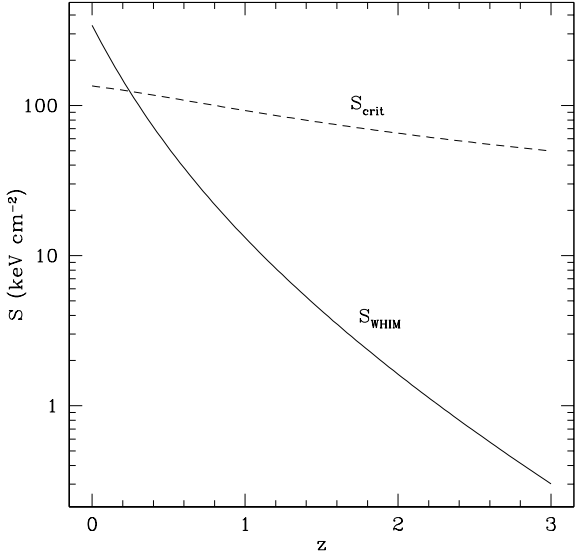,width=80mm}
\caption{The evolution of the median entropy of the WHIM as a function of
  redshift. Note that $Sq_{\rm WHIM} \propto \delta^{1/3}$; here we assume
  $\delta\sim 30$. Although in practice there will be wide scatter
  about this relation, we nonetheless see that the entropy of the WHIM
  falls very rapidly with redshift. In particular, it only exceeds the
  critical entropy $S_{\rm crit}$ required to prevent cooling within a
  Hubble time at late times.}
\label{Fig:S_WHIM}
\end{figure}

We can estimate the redshift evolution of the WHIM
entropy as follows. The postshock temperature of the WHIM can be
estimated as:
\begin{equation}
T_{\rm WHIM}(z) = K (H L_{\rm nl})^{2},
\end{equation}
where $H(z)$ is the Hubble constant, $L_{\rm nl}$ is the nonlinear
length scale in proper coordinates; this produces a good fit to the
results of numerical simulations \cite{CO99,Daveetal}. We have chosen
the normalization constant $K$ so as to reproduce the median
temperature for a given overdensity obtained by \scite{Daveetal} at
z=0. At a given overdensity $\delta$, the entropy of the WHIM scales as
$S_{\rm WHIM} \propto H(z)^{2} L_{\rm nl}^{2} (1+z)^{-2}$; it thus
falls rapidly with redshift. We plot $S_{\rm WHIM}(z)$ in Figure
(\ref{Fig:S_WHIM}); it only exceeds $S_{\rm crit}$ at $z < 1$. In
practice there will be significant scatter about the median entropy of
the WHIM (for instance, the temperature distribution broadens at high
redshift; see Fig. 4 of \pcite{Daveetal}). It would be interesting to
use existing numerical simulations to compute the distribution
function of entropy in the WHIM, as well as the redshift evolution of
the median IGM entropy; such quantities to date have not been
calculated.

It is amusing to speculate on the effect of a cosmological constant on
the future star formation history of our universe. The growth function
is given by \cite{Heath}:
\begin{equation}
D_{1}(a) \propto H(a) \int_{0}^{a} \frac{\dd a^{\prime}}{a^{\prime 3}
  H(a^{\prime})^{3}},
\end{equation} 
where $a=1/(1+z)$ is the scale factor and $H(a)=(\Omega_{\rm m}a^{-3}
+ \Omega_{\rm k}a^{-2} + \Omega_{\Lambda})^{1/2}$, where $\Omega_{\rm
k} = 1- \Omega_{\rm m} - \Omega_{\Lambda}$. Note that at early times
when $a$ is small, $D_{1}(a) \propto a$. If we normalize this so that
$a=1$ and $D_{1}(0)=1$ today, then in the future when $a \rightarrow
\infty$, $D_{1}(a) \rightarrow 1.39$. Thus, structure formation is
already freezing out; the non-linear mass $M_{*}$ will increase by at
most a factor of 4 from its present day value, in contrast to its
rapid growth in the past. Thus, gravitational shock heating is slowing
down and becoming unimportant, as the universe enters a period of
exponential expansion. Gas in already virialized halos will eventually
be able to cool as the universe ages, modulo feedback by star
formation and AGN activity fueled by gas cooling. This is in contrast
to the $\Omega_{\rm m}=1$ SCDM case, when $D_{1}=a$ and structure
formation and gravitational shocks continue apace; in this case most
of the non-stellar baryons in the universe would end up in a hot phase
which will never cool.

\subsection{Could Preheating determine Galaxy Morphology?}

We speculate that preheating may be responsible for the origin of
the Hubble sequence. Gas in dark matter halos which form before the
epoch of preheating are assumed to form an elliptical galaxy in a
rapid collapse, since halos below group scales have gas with cooling
times shorter than their dynamical time $t_{\rm cool} < t_{\rm ff}$
\cite{reesostriker}. On the other hand, halos which collapse after
preheating will accrete gas with entropies more characteristic of
groups and clusters; in particular, they can cool their gas only
slowly, with $t_{\rm cool} > t_{\rm ff}$. Cooling gas in such halos
may instead form a disk galaxy. In fact, \scite{momao} point out that
preheating could potentially solve two problems associated with disk
galaxies: explaining the anomolously low X-ray surface brightnesses
associated with their halos (since preheating decreases gas densities
and X-ray emissivities), and preventing angular momentum transport
from the disk to the dark matter during disk formation (because the
gas distribution is more extended than that of the dark matter, it
retains a higher specific angular momentum during mergers). We note in
passing that the low X-ray luminosities of halos associated with disks
is perhaps unsurprising: a simple extrapolation of the $L_{\rm
X}-T_{\rm X}$ relation to halo temperatures typical for the hosts of
disks ($\sim 0.1$keV) gives X-ray luminosities consistent with
observations (few $\times 10^{40} \, {\rm erg \, s^{-1}}$). Thus,
whatever physics is suppressing the X-ray luminosities of group
operates in disk galaxies too.

\begin{figure}
\psfig{file=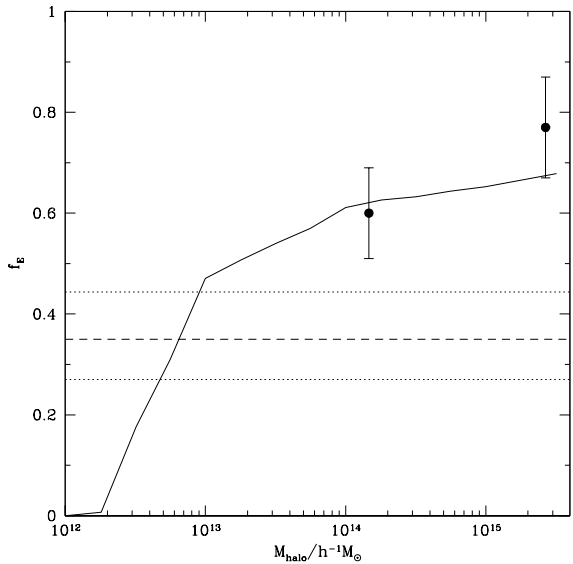,width=80mm}
\caption{The fraction of galaxies formed in $10^{12}M_\odot$ halos
which are ellipticals, as a function of the mass of the halo in which
they are found at $z=0$. Points with errorbars show the determinations
of \protect\scite{balogh02}. (The X-ray luminosities of clusters in
\protect\scite{balogh02} were converted to cluster virial masses using
the relation of \protect\scite{rei99}.) The dashed line shows the
``field'' elliptical fraction from \protect\scite{balogh02}, with
dotted lines showing their errorbars.}
\label{fig:morph}
\end{figure}

In a hierarchical formation scenario, the progenitors of massive
clusters typically formed earlier than those of lower mass
systems. Thus, massive clusters contain a larger fraction of galaxies
which formed prior to the epoch of preheating; the fraction of
elliptical galaxies should increase as the mass of the dark matter
halo increases. To compute this fraction of galaxies which formed
before preheating (and therefore formed ellipticals) we create merger
trees for halos of different masses at $z=0$. We then identify when
$10^{12}M_\odot$ progenitor halos form in these trees. If the
progenitor formed before preheating (assumed to occur at $z=2$ here)
the halo is flagged as hosting an elliptical galaxy, while halos which
collapse later are assumed to host a disk galaxy. In
Figure~\ref{fig:morph} we show the fraction of galaxies formed in
$10^{12}M_\odot$ halos which are ellipticals as a function of the mass
of the halo in which they are found at $z=0$. This is compared against
the observational determinations of \scite{balogh02} of the elliptical
fraction (shown as circles), with their estimate of the ``field''
elliptical fraction shown as a dashed line. Our simple calculation
fits rather well with these data (note that the ``field'' should
correspond approximately to $M_*\sim 10^{13}h^{-1}M_\odot$
halos). Since the typical mass of halos increases in denser
environments we expect this model to produce the observed
morphology-density relation, at least qualitatively. A direct
comparison requires a detailed model of galaxy formation, since it
requires calculating the local galaxy density within the model.

This is of course little more than a consistency check, rather than
proof of principle. The idea that ellipticals collapsed at much higher
redshift than spirals is an old one. It was suggested long ago that
there may be a direct correspondence between peak height and galaxy
morphology, with ellipticals forming from rare ($\sim 3 \sigma$)
density fluctuations, while spirals form from more common ($\sim 2
\sigma$) ones \cite{blumenthal,evrard}; such a scheme can reproduce
their observed relative abundances fairly well and would also explain
the increased clustering strength of ellipticals. Empirically, it is
known that ellipticals exhibit remarkably tight correlations in
various global properties and have very small inferred age dispersion
which indicate a high formation redshift $z \sim 2-3$, unless their
formation was synchronized to an implausible degree. The evidence
includes the tightness of the fundamental plane \cite{vandok}, and the
extreme homogeneity of their optical colors
\cite{boweretal,ellisetal}. However, the physical motivation for early
formation of ellipticals rather than spirals has been rather weak.

The interesting feature that preheating introduces is a physical
reason why there should be a division in morphology between early and
late collapse of halos of the same mass, because of the change in the
cooling properties of the gas. There may be some critical level of
entropy (likely $t_{\rm cool}(S_{\rm crit}) \sim t_{\rm dyn}$) where
galaxy formation switches from one mode to another. To test these
speculations, it would be necessary to conduct high resolution
simulations of a single galaxy, to observe how galaxy morphology and
disk structure changes as the entropy of the accreted gas is
increased. To some extent, we know already from previous simulations
(based on suggestions that photo-ionization could delay cooling until
$z \sim 1$) that feedback processes which delay gas cooling result in
less angular momentum transfer and more disk-like features
\cite{weiletal}.
 
Preheating may also help alleviate possible problems with the dearth
of galaxies found in voids \cite{peebles,mathis,bensonvoids}. Since
the distribution of halo masses in voids is shifted to lower masses
relative to the field, preheating will be particularly effective in
suppressing galaxy formation in these environments. However, lower
mass halos also tend to collapse earlier, so a larger fraction of
these halos will form prior to preheating. This latter effect seems to
dominate. Using dark matter halos from the simulations of
\scite{bensonvoids}, we identified halos which formed after preheating
at $z=2$ (assuming the distribution of formations redshifts given by
\scite{verde} and no correlation of formation time with
environment). Removing these halos from the sample makes only minor
differences to the void probability function for example. Although
these simulations have a rather poor mass resolution for the study of
void galaxies, we expect the effects of preheating to become even less
for lower mass halos, since an even greater fraction of these should
form prior to preheating.

\subsection{Conclusions}

In this paper, we stress that radiative cooling alone cannot account
for the observed entropy floor in clusters and the minimum entropy
required to produce the observed deviation from self-similarity in the
$L_{\rm X}-T_{\rm X}$ relation. This is only possible with severe
overcooling. Thus, some form of entropy injection--perhaps through
supernovae or AGN winds--seems necessary (though radiative cooling
still plays a very important role in reducing the energetic
requirements for producing an entropy floor \cite{voitetal}). The
level of entropy injection required to produce observable changes in
the density profiles of deep potential wells of groups and clusters
must have a drastic effect on smaller halos which host galaxies. We
show that entropy injection can indeed play a very important role in
regulating gas accretion and cooling in galaxies, since gas heated to
the entropy floor can never cool in a Hubble time, regardless of the
densities it is compressed to: for gas temperatures typical of galaxy
halos the cooling time depends almost exclusively on entropy and is
relatively independent of density. Thus, after the epoch of
preheating, the gas supply to galaxies was cut off. This is not
inconsistent with observations if an epoch of preheating at $z \sim 2$
is assumed. We speculate that at a critical level of entropy $t_{\rm
cool}(S_{\rm crit}) \sim t_{\rm ff}$, the mode of gas collapse changes
from free-fall and fragmenation (which produces ellipticals) to
quiescent accretion (which produces spirals).

There are many aspects of the impact of preheating on galaxy formation
which would be worth pursuing in greater detail, particularly with
numerical simulations. One would be the redshift and spatial
dependence of entropy injection--we have only considered a uniform
level of entropy injection which appears instantaneously at $z \sim
2$. Naively, we would expect entropy injection to be proportional to
the integrated star formation, which is proportional to the local
stellar density and local metallicity. Similarly, the spatial
variation and distribution function of entropy due to shock heating of
the IGM deserves attention, and the possibility that the entropy of
the IGM may be responsible for galaxy morphology. It may well be that
the entropy of the IGM regulates the amount of cooling and
star-formation at any given cosmological epoch, and is the chief
mechanism which prevents overcooling at high redshift.

\section*{acknowledgments}

We thank Celine Peroux for providing the data used in Fig
\ref{fig:omega_HI}, and Jerry Ostriker (long ago) and Tommaso Treu for
helpful discussions.

\end{document}